# Sparsity Analysis of a Sonomyographic Muscle-Computer Interface

Nima Akhlaghi[1], Ananya Dhawan[2], Amir A. Khan[1], Biswarup Mukherjee[1], Cecile Truong[1] and Siddhartha Sikdar[1]

[1]Department of Bioengineering & [2]Department of Computer Science, George Mason University

*Abstract*— *Objective: The objectives of this paper are to determine the optimal location for ultrasound transducer placement on the anterior forearm for imaging maximum muscle deformations during different hand motions and to investigate the effect of using a sparse set of ultrasound scanlines for motion classification for ultrasound-based muscle computer interfaces (MCIs). Methods: The optimal placement of the ultrasound transducer along the forearm is identified using freehand 3D reconstructions of the muscle thickness during rest and motion completion. From the ultrasound images acquired from the optimally placed transducer, we determine classification accuracy with equally spaced scanlines across the cross-sectional field-of-view (FOV). Furthermore, we investigated the unique contribution of each scanline to class discrimination using Fisher criteria (FC) and mutual information (MI) with respect to motion discriminability. Results: Experiments with 5 able-bodied subjects show that the maximum muscle deformation occurred between 30-50% of the forearm length for multiple degrees-of-freedom. The average classification accuracy was 94±6% with the entire 128 scanline image and 94±5% with 4 equally-spaced scanlines. However, no significant improvement in classification accuracy was observed with optimal scanline selection using FC and MI. Conclusion: For an optimally placed transducer, a small subset of ultrasound scanlines can be used instead of a full imaging array without sacrificing performance in terms of classification accuracy for multiple degrees-of-freedom. Significance: The selection of a small subset of transducer elements can enable the reduction of computation, and simplification of the instrumentation and power consumption of wearable sonomyographic MCIs particularly for rehabilitation and gesture recognition applications.*

## I. INTRODUCTION

Human machine interfaces that rely on sensing muscle activity have traditionally been a mainstay in assistive technologies for restoring mobility and function, such as prosthetics and exoskeletons [1]–[6]. Recently, muscle computer interfaces (MCIs) have emerged as a new modality for touch-free device-control [7]–[11]. In these scenarios, the aim is the classification of hand gestures or grasps based on noninvasive sensing of muscle activity and interactive control of external devices. In many of these applications, camera-based systems for gesture recognition are not practical. As a result of these advances, there is now renewed interest in developing sophisticated MCI sensing that can accurately classify gestures or grasps.

Surface electromyography (sEMG) - which measures the electrical activity of motor units, is the most popular noninvasive sensing paradigm for MCIs [7]–[11]. The electrical activity of contracting motor units is measured using surface electrodes and the control achieved by mapping these signals to the function of the interactive devices [7], [12], [13]. Particularly for rehabilitation, this technique has become a predominant approach to drive assistive devices such as modern multi-articulated hands [14], [15] and exoskeletons [3], [4]. Recently, gesture recognition systems using sEMG systems have been commercially introduced targeting applications beyond assistive technologies [7]–[11]. However, the sEMG technique suffers from several fundamental limitations such as low signal to noise ratio (SNR) and lack of specificity which lead to limited and non-intuitive control [16]–[18]. On the other hand, pattern recognition approaches using dense sEMG electrode arrays have shown an improvement in prosthetic control [19]–[21]. Although these strategies provide relatively better intuitive control, they do not produce robust graded signals for fine control for applications such as multi-articulated prosthetic hands.

Ultrasound imaging has recently emerged as a new paradigm to sense muscle activity, and a number of studies have shown that it could be an attractive alternative to sEMG-based approaches for MCIs [22]–[29]. The main advantage of the ultrasound-based sensing, or sonomyography, over sEMG is the capability to produce robust signals from contiguous functional compartments deep inside the limb with high specificity. In our previous work, we have demonstrated that ultrasound-based sensing can be used to successfully predict volitional motion intent of both able-bodied and amputee subjects with high accuracy in offline and real-time settings [30]–[32].

Ultrasound imaging systems are increasingly undergoing significant miniaturization and reduction in cost, and portable ultrasound probes that can be connected to and controlled by smartphones are now commercially available [33], [34]. These advancements suggest that ultrasound imaging can be used for developing a wearable sensing system for MCIs. However, traditional ultrasound imaging arrays are bulky and not ideal as a wearable sensor. For a practical ultrasound-based MCI that could be integrated into a compact wearable system, the use of a number of sparsely placed single element ultrasound transducers is more practical than a dense imaging array [29], [35], [36]. Such systems could have a significantly reduced footprint in terms of the transducers, require simpler instrumentation, have reduced computational requirements, as well as reduced power consumption which would be particularly suited for prosthetic control, as well as other gesture recognition applications. However, for these applications there is a need to investigate the effect of scanline (channel) reduction with optimal sensor placement on the motion classification accuracy. To study this effect, this work

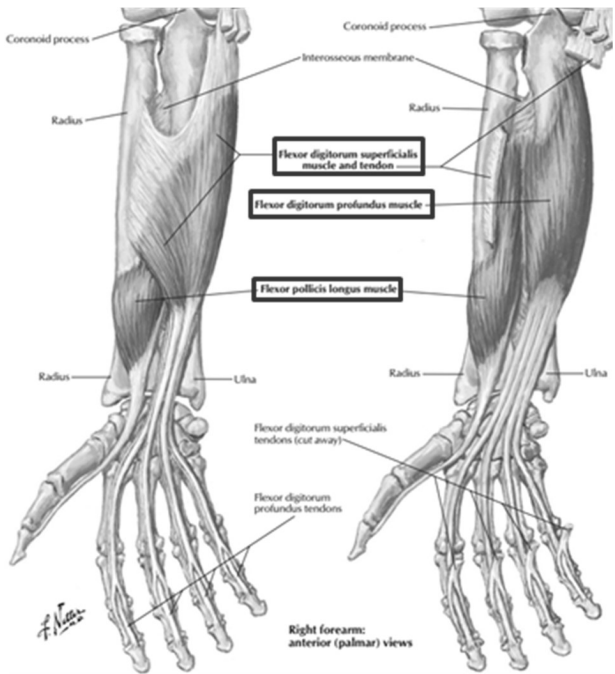

Figure 1. Anatomical drawing of the main flexor muscles of the forearm [43].

had two main objectives:

(1) To determine the optimal transducer position on the anterior forearm to sense the maximum muscle activation during volitional motion intent.

(2) To explore the effect of spatial information reduction on motion classification performance by systematically limiting the number of ultrasound scanlines.

For the first objective, we generated a 3D representation of subject's forearm muscles in rest and motion states (different hand gestures) using freehand imaging technique and identifying the area with maximum muscle activation within the length of the forearm. For the second objective, we placed a clinical ultrasound probe at this optimal position and compare the classification accuracy when different subsets of scanlines. Scanlines were extracted at equally-spaced intervals, or, selected utilizing two widely-used, optimal channel selection strategies. The first method is the distance-based optimal channel subset selection using the Fischer criterion as a distance metric [37], [38]. The second method is the correlation-based optimal channel subset selection using mutual information as a similarity metric [39]–[41] . The following sections describe our methods and results in more details.

## II. Materials and methods

### A. Data acquisition protocols

Ultrasound data were acquired from the forearm of five able-bodied subjects as they performed various tasks. The George Mason University Institutional Review Board approved all study procedures. The experimental procedure was explained to the subjects prior to data collection, and informed consent was obtained from all the subjects.

**Experiment 1: 3D freehand imaging**

For the first experiment, ultrasound imaging data were collected from the forearm of the subjects using a Aixplorer (Supersonic Imagine, Aix-en-Provence, France) ultrasound imaging system with the SL15-4 linear array transducer. The data were acquired from a depth of 4 cm with a frame rate of 50 Hz. The length of each subject's dominant anterior forearm in the supine position was measured from elbow to wrist. An electromagnetic position sensor (3D Guidance TrakStar, Ascension Technologies, VT, USA) was attached to the ultrasound probe to track its location and orientation with respect to a fixed transmitter [42]. The ultrasound probe was placed on the anterior portion of the forearm perpendicular to ulna and freely moved along the forearm length from elbow toward wrist. Each subject was asked to perform four different hand gestures (key grip, pinch grip, power grasp and index pointing) one at the time to activate his/her main forearm flexor muscles: flexor digitorum superficialis (FDS), flexor digitorum profundus (FDP), and flexor pollicis longus (FPL) muscles (Figure 1 [43]). To extract the forearm region with maximum muscle activation, scans were performed at rest (baseline state) and full motion completion for each of the hand gestures. The probe positions, and associated sequential brightness mode images (B-Mode), were streamed into an image analysis software (Stradwin, Cambridge, UK) [42], [44], [45], to reconstruct 3D volumes for further analysis. Figure 2 shows the experimental setup used for freehand ultrasound scans of the forearm.

**Experiment 2: Real-time B-mode ultrasound imaging of forearm muscle.**

For the second experiment, ultrasound imaging data were collected from the forearm of the five subjects using a SonixRP along with L14-5 linear array transducer. The data were acquired from a depth of 4 cm with a frame rate of 50 Hz. The research package of the ultrasound system was used to stream the data directly into MATLAB (The Mathworks Inc., MA, USA), for further processing. The acquired data consisted of 128 beam-formed scanlines of preprocessed RF data. The dynamic range of the envelope of the analytical signal was compressed using the square root function and normalized to reconstruct a B-mode ultrasound image for further analysis. The probe was stabilized on the anterior portion of the subject's forearm perpendicular to the ulna at a proper location based on our finding from 3D freehand imaging analysis, using a custom design cuff. Figure 3 shows the experimental setup for data collection, showing the probe placement as well as the corresponding ultrasound image of cross-sectional anatomy.

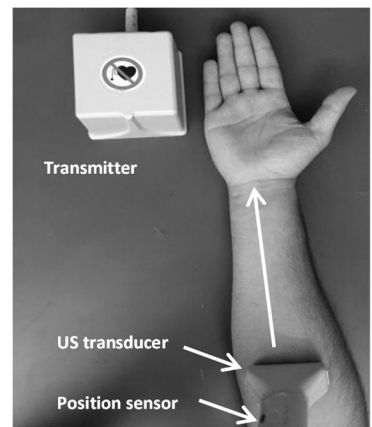

Figure 2. Freehand ultrasound imaging setup. An electromagnetic sensor was attached to the ultrasound probe to track its position (with respect to a fixed transmitter) as it was moved across the forearm.

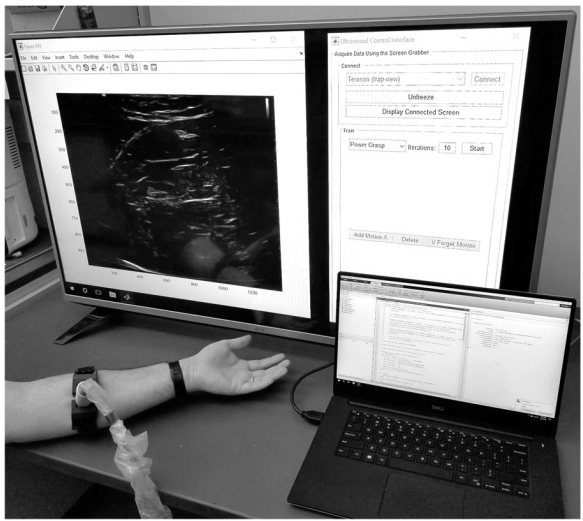

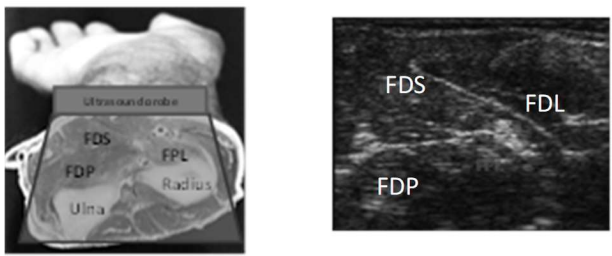

Figure 3. Experimental setup for motion classification data collection (top). Cross-sectional anatomy of the forearm and corresponding B-mode ultrasound image (C). FDS – flexor digitorum superficialis, FDP - flexor digitorum profundus, FPL - flexor pollicis longus

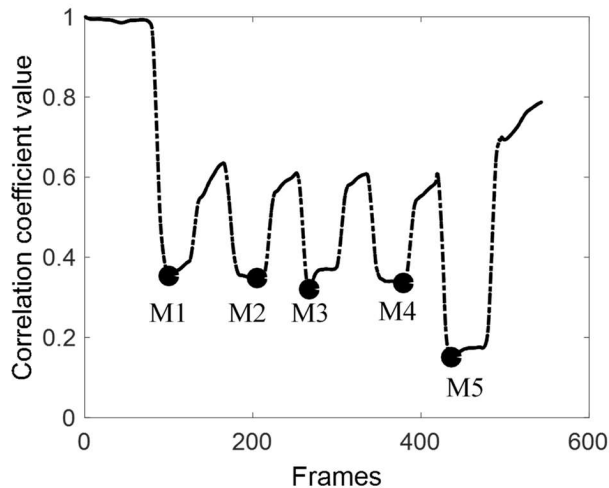

Figure 4. Correlation coefficient for each frame with respect to first frame (considered to be at rest) from an image sequence in one trial. The solid circles correspond to the end-state of each of the 5 performed motions (M1: power grasp, M2: pinch grip, M3: point grip, M4: key grip and M5: wrist pronation).

Each subject was asked to perform 5 consecutive complex hand gestures (power grasp, pinch grip, index pointing, key grip and wrist pronation) on a beep of metronome. Each motion was performed 5 times in a single session. The collected data were further analyzed to investigate the effect of channel selection on the classification accuracy.

### B. Ultrasound data analysis and algorithms:

**Motion classification methodology**

For motion classification, an ultrasound frame corresponding to each motion end-state, i.e. the temporal location of maximum muscle engagement during motion performance, was extracted from collected B-mode ultrasound image sequences over each trial. This was accomplished by calculating the correlation coefficient (CC) between each recorded frames of each trial with respect to the first frame, assumed to be rest-state. The temporal location of each motion end-state frame was then determined as a valley in the calculated CC under the hypothesis that the CC reaches a local minima when the subject has completed a particular motion. These extracted frames served as a feature vector for motion classification. A leave-one-out cross-validation procedure with a Nearest Neighbor classifier was used for classification [31]. The classification results were then evaluated in terms of classification accuracy and represented as a confusion matrix. Figure 4 shows an example of calculated CC value for image sequences of one trial against the rest-state (first frame). The valleys in CC in Figure 4 corresponds to the end-state of each performed motion and the peaks in the CC correspond to the rest-state. The variability in the peak value can be attributed to the fact that the subjects were unable to return to the same rest-state consistently.

**Scanline reduction approaches**

We then considered three approaches of scanline reduction and investigated its effect on motion classification accuracy. Low-dimensional feature maps were reconstructed using an extracted subset of the full 128 scanlines and classification accuracy results were compared against the original data. For the first approach, we select different subsets of uniformly distributed scanlines from the full 128-scanline, B-mode images. Further, we adapt two widely used channel selection strategies from BCI and sEMG literature [46]–[52] to systematically select subset of scanlines based on their contribution in class discrimination. Implementation details of these approaches are described below.

*a. Uniformly distributed scanline selection (UDSS)*

In this method, 4, 8 and 16 subsets of equally spaced scanlines across the ultrasound field of view (FOV) were selected from 128 scanlines under the assumption that all scanlines have similar contribution to class discriminability. A low-dimensional feature map was then reconstructed by stacking the extracted scanlines in a down-sampled 2D representation.

### b. Distance-based scanline selection (DSS) using Fisher Criterion

In this approach, we use Fisher criterion (FC) as a distance measure to determine the class discrimination impact of each ultrasound scanline. FC maximizes the inter-class separation while minimizing the intra-class variance. Given two examples feature vectors $C_1$ and $C_2$ of classes one and two respectively, the FC score for $j^{th}$ channel (ultrasound scanline) can be calculated as:

$$R_j(C_1, C_2) = \frac{(\mu_j(C_1) - \mu_j(C_2))^2}{V_j(C_1) + V_j(C_2)} \quad (1)$$

where $\mu$ is the mean and $V$ is the variance of the corresponding example feature vector.

```
Input: Original data: D_{n×m×d×k}
Output: FC/MI scores: FC_{m×n} /MI_{m×n}
n : number of classes
d : depth samples
m : number of scanlines (channels) per class
k : number of trial per class

foreach class i
    foreach trial t
        foreach channel j
            R^i(j, 1:n-1, t) ← (μ_j(C_i) - μ_j(C_{1:n-1}))^2 / (V_j(C_i) + V_j(C_{1:n-1}))
            I^i(j, 1:n-1, t) ← H(C_i) + H(C_{1:n-1}) - H(C_i; C_{1:n-1})
        end
        Rsum^i(j, t) ← Σ_{n-1} R_{m×n-1×t}
        Isum^i(j, t) ← Σ_{n-1} I_{m×n-1×t}
    end
    FC (i, j) ← average Rsum^i_{m×k} across k trials
    MI (i, j) ← average Rsum^i_{m×k} across k trials
end
Output : FC_{n×m} and MI_{n×m}
```

Figure 5. Pseudocode of Fischer criterion (FC) and Mutual information (MI) calculation algorithms.

Here we used a generalized one-vs-all version of DSS for multiclass analysis in which each individual class is compared against all the other classes. Using eq. (1), a score is calculated for each scanline in a given class and the corresponding scanlines in the remaining classes, i.e., given $n$ classes and $m$ scanlines we compute $m \times n-1$ scores for each class. The computed scores for a given class are then summed together resulting in a single value for each of the $m$ scanlines. This value represents the contribution of that scanline in the discriminability of a class against all other classes. This calculation is repeated for all trials such that given $k$ trials each class is represented by $m \times k$ values. For every class, the computed $m \times k$ matrix is then averaged across the $k$ trials and normalized by the maximum FC score of that class. This results in an $m \times n$ matrix of normalized FC scores for $n$ classes.

### c. Correlation-based scanline selection (CSS) using Mutual information

In this method, we utilize mutual information (MI) as a measure of similarity to rank and select a subset of scanlines. Since MI computes similarity between two classes, scanlines with low MI values (low similarity across classes) are desirable as they provide higher discriminability. Given two feature vectors $C_1$ and $C_2$ of classes one and two respectively, the MI score for $j^{th}$ scanline can be defined as:

$$I_j(C_1, C_2) = H_j(C_1) + H_j(C_2) - H_j(C_1; C_2) \quad (2)$$

Where the $H_j(C_1)$ and $H_j(C_2)$ represent the entropies of $C_1$ and $C_2$ for the $j^{th}$ scanline respectively and $H_j(C_1;C_2)$ represents the joint entropy of the two feature vectors for the $j^{th}$ scanline. These entropies can be defined as follows:

$$H_j(C_1) = -\sum_{C_1} p(C_1) \log p(C_1) \quad (3)$$

$$H_j(C_2) = -\sum_{C_2} p(C_2) \log p(C_2) \quad (4)$$

$$H_j(C_1; C_2) = -\sum_{C_1, C_2} p(C_1, C_2) \log p(C_1, C_2) \quad (5)$$

Where, $p$ represents the probability mass function for a particular class. A generalized implementation of CSS similar to that of DSS for multiclass analysis is used here. Figure 5 illustrates the pseudocode implementation of both DSS and CSS.

## III. RESULTS

### A. Location of maximum muscle activation on the forearm

Figure 6 shows the results from 3D ultrasound analysis to find the location of maximum muscle activity averaged across all subjects when performing complex hand motions. These results demonstrate that the maximum activation of the main flexor muscles (FDS, FDP, and FPL) occurs approximately in the range of 40% to 50% of the forearm length from the elbow joint and therefore can serve as the optimal zone for ultrasound probe placement. The variability in location of maximum muscle activations can be attributed to anatomical differences between subjects. For subsequent experiments, the probe position was first positioned in this optimal zone and then visually reoriented to ensure that all the flexor muscles are within the ultrasound FOV.

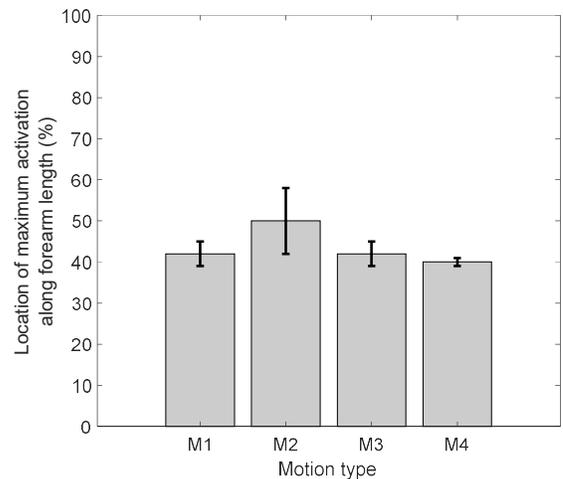

Figure 6. Average location in percentage of length of the forearm (from elbow) with maximum muscle deformation during complex hand motions. (M1-key grip, M2-Pinch, M3-Index point, M4-Power grasp).

### B. Comparison of scanline selection strategies

For each subject, data were then collected for 5 different motions and FC and MI score matrices were computed in order to extract the location of scanlines with highest class discriminability. This is done by summing the scores across classes so that the total discriminability of a given scanline for all classes can be computed. Since higher FC scores result in better discrimination between classes, we identify the positions of the local maxima in the aggregated signal for the DSS approach, and sort them in a descending manner. However, since lower MI values correspond to higher discriminability, we identify the positions of local minima in the aggregated signal and sort them in ascending order. 4, 8 and 16 of the highest discriminability scanline subsets were selected based on the FC and MI scores and the classification performance was evaluated for each case.

Figure 7A shows the *m x n (m=5, n = 128)*, FC score matrix for one subject as a function of scanline and motion class. Higher intensities in this figure represent higher discrimination capability of the corresponding scanline for a given motion class. Figure 7B illustrates the aggregation of the FC scores across all classes to obtain the normalized score and identified extrema (local maxima) indicate the most discriminative spatial locations for a given subject. Similarly, Figure 8A shows the *m x n*, MI score matrix for the same subject. Unlike FC scores (Figure 7A) however; brighter intensities in Figure 9A represent lower discriminability. Figure 8B illustrates the aggregation of the MI scores across all classes and the identified extrema (local minima) indicate the most discriminative spatial locations for a given subject. These locations are independently sorted in the order of their discrimination capability.

The inter-trial consistency of FC and MI scores in terms of average standard deviation across all trial for each subject is provided in Table I. The same variability is also shown in Figures 9A and 9B. These results demonstrate that the extracted aggregated scores for each trail have similar patterns and that the discriminative scanline locations appear in the same spatial region. Therefore, it is justified to extract optimal scanline based on the average signal across trials.

TABLE I. Average standard deviation of FC and MI scores across five trials and two approaches for each subject.

| Subject | 1 | 2 | 3 | 4 | 5 |
|---|---|---|---|---|---|
| DSS | 0.06 | 0.06 | 0.1 | 0.05 | 0.06 |
| CSS | 0.02 | 0.02 | 0.02 | 0.03 | 0.02 |

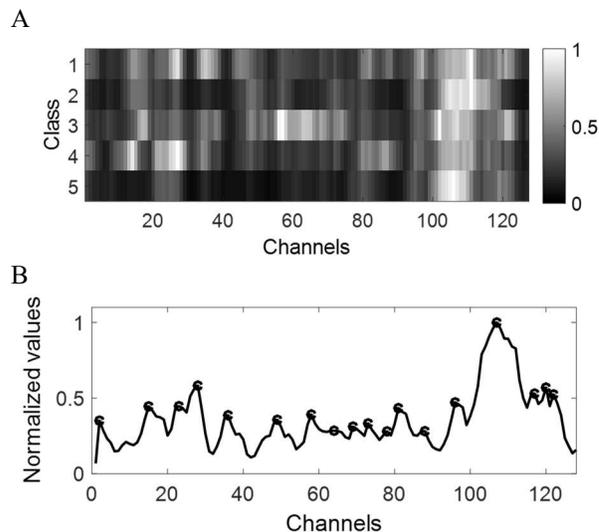

Figure 7. FC score map (A), representing each channel's relative capacity to distinguish a given motion from other motions. The normalized sum of FC scores across all classes (B), where the peak locations correspond to the optimal channel offering maximum motion class discriminability.

The motion classification accuracy using nearest neighbor-based leave-one-out cross-validation technique was then computed for all three scanline reduction approaches and are shown in Table II. These results represent the effect of the scanline reduction on the CA using three proposed methods averaged over all motion for each subject. For each of the three techniques the CA was calculated for three different subsets (4, 8, and 16 scanlines) and compared to those obtained using feature map from the original 128 scanline data. From Table II it can be observed that the scanline reduction using any of the proposed methods does not have any significant effect on the CA in compare to those obtained using full 128 scanline. It can be also noted that the use of uniformly distributed scanlines provides comparable CA results to those calculated using the adopted optimal channel selection approaches (DSS, CSS).

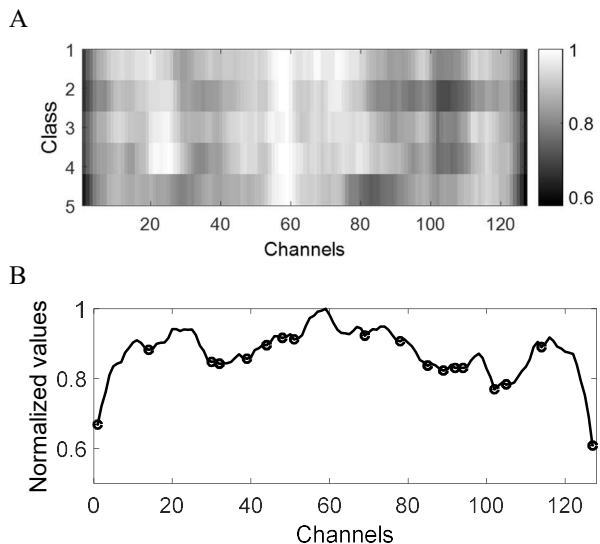

Figure 8. Mutual information score map (A), calculated as a measure of channels separation impact for each individual class. The normalized sum of MI scores across all classes (B), where the valley locations correspond to the optimal channel for maximum class separation across all different motion classes.

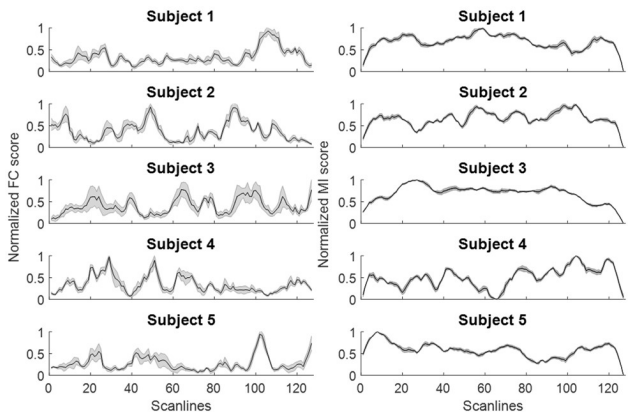

Figure 9. Aggregated Fischer criterion (FC) and mutual information (MI) scores across all classes for each individual trials for all subjects. The solid black line represents the average signal across all trials with the standard deviation denoted by the shaded region.

## IV. DISCUSSION

In this paper we investigated the feasibility of using a sparse set of ultrasound scanlines for classification of hand motions using sonomyography. We first investigated optimal sensor placement on the anterior forearm and then investigated the effect of scanline reduction on classification accuracy.

Our results have shown that the largest deformation (change in muscle thickness) of the forearm flexors appears between 40 to 50% of the forearm length from the elbow during the performance of a range of complex hand motions. This observation corroborates well with known functional anatomy of the three primary forearm flexors, FDS, FDP and FPL [53]. The highest muscle deformation is likely to occur within the first half of the forearm, where the muscle belly of the three primary flexors is located (Figure 1). Ultrasound imaging allows us to reproducibly identify the appropriate region of interest in able bodied subjects with a relatively low inter-subject variability. For amputee subjects, the functional anatomy is expected to be unique for each individual due to the differences in level of amputation and the surgical intervention performed (myodesis vs myoplasty [54]). Ultrasound imaging can potentially enable a determination of the functional anatomy for each amputee for personalized sensor placement for reliable performance in ultrasound-based MCIs. We anticipate that this ultrasound imaging method can also be used for optimal placement of sEMG electrodes; however, this needs to be investigated in the future.

We then explored the effect of limiting the number of ultrasound scanlines on classification performance within this optimal zone on the forearm. The main motivation for scanline selection is to investigate whether it is possible to utilize a small number of single-element ultrasound transducers rather than imaging arrays while maintaining reliable performance [55].

We compared a simple equidistant scanline distribution approach to adapted versions of a distance-based (DSS) and a correlation-based (CSS) feature/channel subset selection strategies commonly used by the brain-computer interface and sEMG community [46]–[52]. DSS evaluates the class discrimination impact of each scanline using a distance measure such as Fisher's criterion (FC). On the other hand, CSS uses mutual information (MI) between the respective scanlines across classes.

Our results show that equidistant scanline distribution yields classification performance that is comparable to discriminative scanline selection with FC or MI. For all three scanline reduction methodologies, our cross-validation accuracy results demonstrate that the number of scanlines can be reduced from 128 to 4 without significant effect on the performance. Given all 128 scanlines, the average CA across five motion classes and five subjects, was 94±6 %. With 4 selected scanlines, the average CA was 94±5%, 94±8 % and 94±6 % for equidistant scanline distribution, FC and MI respectively across subjects. As shown in Figure 9, the FC and MI signatures were consistent over trials. It is important to note however, that the computed signature is subject-specific, as is expected due to the slight differences in anatomical structure and probe placement.

Although using four scanlines was sufficient for maintaining the classification performance, generally, including more data scanlines resulted in a slight improvement in CA. The average CA for FC was 94±10 % and 95±4% for 8 and 16 scanlines respectively. Likewise, the CA for MI was 95±3 % and 97±3% for 8 and 16 scanlines respectively. These results show that the ultrasound data being rich in spatial information content, a sparse, equally-spaced transducer configuration that images all the principal forearm muscle groups is sufficient for complex hand motion classification applications.

TABLE II. Comparison of classification accuracy for subsets of scanlines selected using UDSS, DSS and CSS techniques.

| | UDSS (Uniformly-spaced) | | | DSS (FC) | | | CSS (MI) | | | Original data |
|---|---|---|---|---|---|---|---|---|---|---|
| **Number of scanlines** | 4 | 8 | 16 | 4 | 8 | 16 | 4 | 8 | 16 | 128 |
| **Subject 1** | 100% | 100% | 100% | 100% | 100% | 100% | 100% | 100% | 100% | 100% |
| **Subject 2** | 88% | 88% | 88% | 80% | 76% | 88% | 88% | 92% | 96% | 92% |
| **Subject3** | 96% | 100% | 96% | 100% | 100% | 96% | 92% | 96% | 96% | 96% |
| **Subject 4** | 92% | 96% | 96% | 96% | 100% | 96% | 100% | 96% | 100% | 96% |
| **Subject 5** | 92% | 92% | 92% | 96% | 96% | 96% | 88% | 92% | 92% | 84% |
| **Average** | **94±5%** | **95±4%** | **94±4%** | **94±8%** | **94±10 %** | **95±4 %** | **94±6 %** | **95±3%** | **97±3%** | **94±6%** |

This work has been submitted to the IEEE for possible publication. Copyright may be transferred without notice, after which this version may no longer be accessible.7Additional studies are needed to investigate the anatomical dependence of optimal channel selection by determining whether the consistency in the locations identified through DSS and CSS extend to other more complex channel selection paradigms, such as common spatial patterns [55].

The ability to classify motions with high accuracy using small number of transducers is critical in prosthetic control applications and other applications because it enables optimal system design on an individual basis such that the operation of given device is appropriately tuned to user behavior. Such design permits more intuitive control of device functionality and thus is likely to reduce user frustration. Furthermore, the ability to reduce the number of required transducers and associated instrumentation, without sacrificing the classification accuracy, increases the practicality of a gesture recognition system by reducing both the hardware complexity and the power consumption requirements, desirable traits for a practical MCI.

## V. Conclusion

In conclusion, this paper describes a systematic procedure to quantify the number and appropriate placement of ultrasound transducers required for control of MCIs. Our results show that only a small subset of scanlines can be used instead of a full imaging array without significant degradation of classification accuracy. In future, we hope to utilize the techniques detailed in this paper to design practical wearable systems and quantify its performance in order to further validate our hypothesis that a small number of ultrasound transducers can be used to achieve high gesture recognition accuracy.

## References

[1] M. Zecca, S. Micera, M. C. Carrozza, and P. Dario, "Control of Multifunctional Prosthetic Hands by Processing the Electromyographic Signal," *Crit. Rev. Biomed. Eng.*, vol. 30, no. 4–6, pp. 459–485, 2002.

[2] C. Cipriani, F. Zaccone, S. Micera, and M. Chiara Carrozza, "On the Shared Control of an EMG-Controlled Prosthetic Hand: Analysis of User–Prosthesis Interaction," *IEEE Trans. Robot.*, vol. 24, no. 1, 2008.

[3] K. Kiguchi and Y. Hayashi, "An EMG-Based Control for an Upper-Limb Power-Assist Exoskeleton Robot," *IEEE Trans. Syst. Man, Cybern. Part B*, vol. 42, no. 4, pp. 1064–1071, Aug. 2012.

[4] T. Lenzi, S. M. M. De Rossi, N. Vitiello, and M. C. Carrozza, "Intention-Based EMG Control for Powered Exoskeletons," *IEEE Trans. Biomed. Eng.*, vol. 59, no. 8, pp. 2180–2190, Aug. 2012.

[5] P. Geethanjali, "Myoelectric control of prosthetic hands: state-of-the-art review.," *Med. devices*, vol. 9, pp. 247–55, 2016.

[6] A. Fougner, O. Stavdahl, P. J. Kyberd, Y. G. Losier, and P. A. Parker, "Control of Upper Limb Prostheses: Terminology and Proportional Myoelectric Control—A Review," *IEEE Trans. Neural Syst. Rehabil. Eng.*, vol. 20, no. 5, pp. 663–677, Sep. 2012.

[7] K. Tanaka, K. Matsunaga, and H. O. Wang, "Electroencephalogram-based control of an electric wheelchair," *IEEE Trans. Robot.*, vol. 21, no. 4, pp. 762–766, Aug. 2005.

[8] T. S. Saponas, D. S. Tan, D. Morris, R. Balakrishnan, J. Turner, and J. A. Landay, "Enabling Always-Available Input with Muscle-Computer Interfaces," in *Proceedings of the 22Nd Annual ACM Symposium on User Interface Software and Technology*, 2009, pp. 167–176.

[9] T. S. Saponas, D. S. Tan, D. Morris, J. Turner, and J. A. Landay, "Making Muscle-Computer Interfaces More Practical," in *Proceedings of the SIGCHI Conference on Human Factors in Computing Systems*, 2010, pp. 851–854.

[10] H. Benko, T. S. Saponas, D. Morris, and D. Tan, "Enhancing input on and above the interactive surface with muscle sensing," in *Proceedings of the ACM International Conference on Interactive Tabletops and Surfaces - ITS '09*, 2009, p. 93.

[11] T. S. Saponas, D. S. Tan, D. Morris, and R. Balakrishnan, "Demonstrating the Feasibility of Using Forearm Electromyography for Muscle-Computer Interfaces," in *Proceedings of the SIGCHI Conference on Human Factors in Computing Systems*, 2008, pp. 515–524.

[12] R. Song, K. Y. Tong, X. Hu, and L. Li, "Assistive control system using continuous myoelectric signal in robot-aided arm training for patients after stroke," *IEEE Trans. Neural Syst. Rehabil. Eng.*, vol. 16, no. 4, pp. 371–379, Aug. 2008.

[13] S. A. Dalley, H. A. Varol, and M. Goldfarb, "A method for the control of multigrasp myoelectric prosthetic hands," *IEEE Trans. Neural Syst. Rehabil. Eng.*, vol. 20, no. 1, pp. 58–67, Jan. 2012.

[14] C. Castellini and P. Van Der Smagt, "Surface EMG in advanced hand prosthetics," *Biol. Cybern.*, vol. 100, no. 1, pp. 35–47, 2009.

[15] A. H. Al-Timemy, G. Bugmann, J. Escudero, and N. Outram, "Classification of finger movements for the dexterous hand prosthesis control with surface electromyography," *IEEE J. Biomed. Heal. Informatics*, vol. 17, no. 3, pp. 608–618, May 2013.

[16] H. S. Ryait, A. S. Arora, and R. Agarwal, "Study of issues in the development of surface EMG controlled human hand," in *Journal of Materials Science: Materials in Medicine*, 2009, vol. 20, no. SUPPL. 1, pp. 107–114.

[17] R. Vinjamuri, Z.-H. H. Mao, R. Sclabassi, and M. Sun, "Limitations of Surface EMG Signals of Extrinsic Muscles in Predicting Postures of Human Hand," in *2006 International Conference of the IEEE Engineering in Medicine and Biology Society*, 2006, vol. 1, pp. 5491–5494.

[18] C. Disselhorst-Klug, T. Schmitz-Rode, and G. Rau, "Surface electromyography and muscle force: Limits in sEMG-force relationship and new approaches for applications," *Clinical Biomechanics*, vol. 24, no. 3. pp. 225–235, Mar-2009.

[19] A. B. Ajiboye and R. F. F. Weir, "A heuristic fuzzy logic approach to EMG pattern recognition for multifunctional prosthesis control," *IEEE Trans. Neural Syst. Rehabil. Eng.*, vol. 13, no. 3, pp. 280–291, Sep. 2005.

[20] J. U. Chu, I. Moon, and M. S. Mun, "A real-time EMG pattern recognition system based on linear-nonlinear feature projection for a multifunction myoelectric hand," *IEEE Trans. Biomed. Eng.*, vol. 53, no. 11, pp. 2232–2239, Nov. 2006.

[21] G. Li, A. E. Schultz, and T. A. Kuiken, "Quantifying pattern recognition- based myoelectric control of multifunctional transradial prostheses," *IEEE Trans. Neural Syst. Rehabil. Eng.*, vol. 18, no. 2, pp. 185–192, Apr. 2010.

[22] X. Chen, Y. P. Zheng, J. Y. Guo, and J. Shi, "Sonomyography (smg) control for powered prosthetic hand: A Study with normal subjects," *Ultrasound Med. Biol.*, vol. 36, no. 7, pp. 1076–1088, Jul. 2010.

[23] Y. P. Zheng, M. M. F. F. Chan, J. Shi, X. Chen, and Q. H. Huang, "Sonomyography: Monitoring morphological changes of forearm muscles in actions with the feasibility for the control of powered prosthesis," *Med. Eng. Phys.*, vol. 28, no. 5, pp. 405–415, Jun. 2006.

[24] J. Shi, J.-Y. Y. Guo, S.-X. X. Hu, and Y.-P. P. Zheng, "Recognition of finger flexion motion from ultrasound image: a feasibility study," *Ultrasound Med. Biol.*, vol. 38, no. 10, pp. 1695–1704, Oct. 2012.

[25] J. Shi, S. Hu, Z. Liu, J.-Y. Guo, Y. Zhou, and Y. Zheng, "Recognition of Finger Flexion from Ultrasound Image with Optical Flow: A Preliminary Study," in *2010 International Conference on Biomedical Engineering and Computer Science*, 2010, pp. 1–4.

[26] J. Shi, Q. Chang, and Y.-P. Zheng, "Feasibility of controlling prosthetic hand using sonomyography signal in real time: Preliminary study," *J. Rehabil. Res. Dev.*, vol. 47, no. 2, pp. 87–98, 2010.

[27] C. Castellini, K. Hertkorn, M. Sagardia, D. S. Gonzalez, and M. Nowak, "A virtual piano-playing environment for rehabilitation based upon ultrasound imaging," in *5th IEEE RAS/EMBS International Conference on Biomedical Robotics and Biomechatronics*, 2014, pp. 548–554.

[28] C. Castellini, G. Passig, and E. Zarka, "Using ultrasound images of the forearm to predict finger positions," *IEEE Trans. Neural Syst. Rehabil. Eng.*, vol. 20, no. 6, pp. 788–797, Nov. 2012.

[29] J. McIntosh, A. Marzo, M. Fraser, and C. Phillips, "EchoFlex," in *Proceedings of the 2017 CHI Conference on Human Factors in Computing Systems - CHI '17*, 2017, pp. 1923–1934.

[30] S. Sikdar *et al.*, "Novel method for predicting dexterous individual finger movements by imaging muscle activity using a wearable

ultrasonic system," *IEEE Trans. Neural Syst. Rehabil. Eng.*, vol. 22, no. 1, pp. 69–76, Jan. 2014.

[31] N. Akhlaghi *et al.*, "Real-time classification of hand motions using ultrasound imaging of forearm muscles," *IEEE Trans. Biomed. Eng.*, vol. 63, no. 8, pp. 1687–1698, Aug. 2016.

[32] C. A. Baker, N. Akhlaghi, H. Rangwala, J. Kosecka, and S. Sikdar, "Real-time, ultrasound-based control of a virtual hand by a trans-radial amputee," *Proc. Annu. Int. Conf. IEEE Eng. Med. Biol. Soc. EMBS*, vol. 2016–Octob, pp. 3219–3222, Aug. 2016.

[33] W. D. Richard, D. M. Zar, and R. Solek, "A Low-Cost B-Mode USB Ultrasound Probe," *Ultrason. Imaging*, vol. 30, no. 1, pp. 21–28, Jan. 2008.

[34] Philips, "Philips Lumify, portable ultrasound," 2016. [Online]. Available: https://www.lumify.philips.com/web/?&origin=%7Cmckv%7Cs_dc&pcrid=240894965887%7Cplid%7C. [Accessed: 24-Jan-2018].

[35] Y. Li, K. He, X. Sun, and H. Liu, "Human-machine interface based on multi-channel single-element ultrasound transducers: A preliminary study," in *2016 IEEE 18th International Conference on e-Health Networking, Applications and Services, Healthcom 2016*, 2016, pp. 1–6.

[36] N. Hettiarachchi, Z. Ju, and H. Liu, "A New Wearable Ultrasound Muscle Activity Sensing System for Dexterous Prosthetic Control," in *2015 IEEE International Conference on Systems, Man, and Cybernetics*, 2015, pp. 1415–1420.

[37] R. N. Khushaba and A. Al-Jumaily, "Channel and feature selection in multifunction myoelectric control," in *Conference proceedings: ... Annual International Conference of the IEEE Engineering in Medicine and Biology Society. IEEE Engineering in Medicine and Biology Society. Annual Conference*, 2007, vol. 2007, pp. 5182–5185.

[38] H. M. Al-Angari, G. Kanitz, S. Tarantino, J. Rigosa, and C. Cipriani, "Feature and Channel Selection Using Correlation Based Method for Hand Posture Classification in Multiple Arm Positions," in *Replace, Repair, Restore, Relieve--Bridging Clinical and Engineering Solutions in Neurorehabilitation*, Springer, Cham, 2014, pp. 227–236.

[39] H. M. Al-Angari, G. Kanitz, S. Tarantino, and C. Cipriani, "Distance and mutual information methods for EMG feature and channel subset selection for classification of hand movements," *Biomed. Signal Process. Control*, vol. 27, pp. 24–31, May 2016.

[40] M. Arvaneh, C. Guan, K. K. Ang, and H. C. Quek, "EEG Channel Selection Using Decision Tree in Brain-Computer Interface," *apsipa.org*, no. December, pp. 225–230, 2010.

[41] H. Tang, H. Maitre, N. Boujemaa, and W. Jiang, "On the relevance of linear discriminative features," *Inf. Sci. (Ny).*, vol. 180, no. 18, pp. 3422–3433, Sep. 2010.

[42] G. M. Treece, A. H. Gee, R. W. Prager, C. J. . Cash, and L. H. Berman, "High-definition freehand 3-D ultrasound," *Ultrasound Med. Biol.*, vol. 29, no. 4, pp. 529–546, Apr. 2003.

[43] F. H. Netter, *Atlas of Human Anatomy*. 2006.

[44] A. Gee, R. Prager, G. Treece, C. Cash, and L. Berman, "Processing and visualizing three-dimensional ultrasound data," *Br. J. Radiol.*, vol. 77, no. suppl_2, pp. S186--S193, Dec. 2004.

[45] G. Treece, R. Prager, A. Gee, and L. Berman, "3D ultrasound measurement of large organ volume.," *Med. Image Anal.*, vol. 5, no. 1, pp. 41–54, Mar. 2001.

[46] A. Al-Ani and A. Al-Sukker, "Effect of feature and channel selection on EEG classification," in *Annual International Conference of the IEEE Engineering in Medicine and Biology - Proceedings*, 2006, pp. 2171–2174.

[47] T. Shibanoki, K. Shima, T. Tsuji, A. Otsuka, and T. Chin, "A quasi-optimal channel selection method for bioelectric signal classification using a partial Kullback-Leibler information measure," *IEEE Trans. Biomed. Eng.*, vol. 60, no. 3, pp. 853–861, Mar. 2013.

[48] Y. Geng, X. Zhang, Y.-T. Zhang, and G. Li, "A novel channel selection method for multiple motion classification using high-density electromyography," *Biomed. Eng. Online*, vol. 13, no. 1, p. 102, Jul. 2014.

[49] H. J. Hwang, J. M. Hahne, and K. R. Müller, "Channel selection for simultaneous myoelectric prosthesis control," in *2014 International Winter Workshop on Brain-Computer Interface, BCI 2014*, 2014, pp. 1–4.

[50] J. P. W. Pluim, J. B. A. Maintz, and M. A. Viergever, "Mutual-information-based registration of medical images: a survey," *IEEE Trans. Med. Imaging*, vol. 22, no. 8, pp. 986–1004, Aug. 2003.

[51] F. Lotte and C. Guan, "Regularizing common spatial patterns to improve BCI designs: Unified theory and new algorithms," *IEEE Trans. Biomed. Eng.*, vol. 58, no. 2, pp. 355–362, Feb. 2011.

[52] K. P. Thomas, C. T. Lau, A. P. Vinod, C. Guan, and K. K. Ang, "A New Discriminative Common Spatial Pattern Method for Motor Imagery Brain—Computer Interfaces," *IEEE Trans. Biomed. Eng.*, vol. 56, no. 11, pp. 2730–2733, Nov. 2009.

[53] J. A. Gosling, *Atlas of human anatomy with integrated text*. J.B. Lippincott Co., 1985.

[54] L. S. M. Tintle, L. M. F. Baechler, C. G. P. Nanos, L. J. A. Forsberg, and M. B. K. Potter, "Traumatic and Trauma-Related Amputations," *J. Bone Jt. Surgery-American Vol.*, vol. 92, no. 18, pp. 2934–2945, Dec. 2010.

[55] T. Alotaiby, F. E. A. El-Samie, S. A. Alshebeili, and I. Ahmad, "A review of channel selection algorithms for EEG signal processing," *EURASIP J. Adv. Signal Process.*, vol. 2015, no. 1, p. 66, Dec. 2015.